\documentclass{article}
\usepackage{amsmath,amssymb}
\newcommand{\be}{\begin{equation}}
\newcommand{\ee}{\end{equation}}
\newcommand{\bea}{\begin{eqnarray}}
\newcommand{\eea}{\end{eqnarray}}
\newcommand{\bean}{\begin{eqnarray*}}
\newcommand{\eean}{\end{eqnarray*}}

\def\beq{\begin{equation}}

\def\eeq{\end{equation}}
\def\R{\mathcal{R}}

\relax

\def\d{\partial}

\def\a{\alpha'}

\begin{document}

\title{Dilatonic black holes in superstring gravity}

\author{Filipe Moura
\\
Departamento de Matem\'atica, Escola de Tecnologias e Arquitetura and \\Instituto de Telecomunica\c c\~oes, \\ISCTE - Instituto Universit\'ario de Lisboa,
\\Av. das For\c cas Armadas, 1649-026 Lisboa, Portugal\\
\\
\texttt{fmoura@lx.it.pt}
}
\maketitle

\begin{abstract}
We solve the dilaton field equation in the background of a spherically symmetric black hole in type II superstring theory with $\a^3$ corrections in arbitrary $d$ spacetime dimensions. We then apply this result to obtain a spherically symmetric black hole solution with $\a^3$ corrections, in superstring theory compactified on a torus, coupled to such dilaton. For this black hole we obtain its mass, entropy, temperature, specific heat and free energy.
\end{abstract}

\newpage

\section{Introduction}
\noindent

A frequently considered effect of string theory is the result of corrections in the inverse string tension $\a$ in the form of higher derivative terms in the effective action. Naturally, such corrections are also manifest in solutions to the field equations from such actions.

String $\a$ corrections to black hole solutions has been a very active topic of research, in different dimensions and for many kinds of black holes. Concerning spherically symmetric $d-$dimensional black holes, curvature--squared corrections (first order in $\a$) were first obtained in \cite{cmp89}; those were the leading corrections in bosonic and heterotic string theories. Higher order corrections, quartic in the Riemann tensor (third order in $\a$) to the same black holes were obtained in \cite{Myers:1987qx}; these are the leading corrections in type II superstring theories.

In article \cite{Moura:2009it} we studied the effects of string compactification on a torus from 10 (or 26) to arbitrary $d$ dimensions on spherically symmetric black holes with corrections of order $\a$, as those considered in \cite{cmp89}. In this article, we extend such study to black holes of the same type but with $\a^3$ corrections, as those considered in \cite{Myers:1987qx}.

These solutions may be written both in the string and Einstein frames. In order to pass from one frame to another, one must perform a conformal transformation involving the dilaton field. Therefore in our case we should determine the solution to the dilaton in the background of an $\a^3$-corrected spherically symmetric black hole.

The article is organized as follows: in section 2, we will solve the dilaton field equation in $d$ dimensions, in the background of a spherically
symmetric black hole, in the presence of curvature corrections of order $\a^3$. In section 3 we revise the $\a^3-$corrected noncompactified solution of \cite{Myers:1987qx}: we obtain some of its thermodynamical properties (mass, entropy, temperature, specific heat and free energy); a few of which had never been previously computed. In section 4 we present the calculations leading to the $\a^3-$corrected $d-$dimensional solution compactified on a torus. Finally, in section 5 we re-derive the same thermodynamical properties for the black hole solution obtained in section 4, and we compare with the results of section 3, evaluating the effects of the compactification.

\section{The dilaton in the background of a $d-$dimensional black hole with $\a^3$ corrections}
\noindent

The most general static, spherically symmetric metric in $d$ spacetime
dimensions can be written in spherical coordinates as
\be
d\,s^2=-f(r)\,d\,t^2 + g^{-1}(r)\,d\,r^2 +r^2\,d\,\Omega^2_{d-2}.
\label{metric1} \ee $f, g$ are arbitrary functions of the radius
$r;$ $d\,\Omega^2_{d-2}=\sum_{i=2}^{d-1} \prod_{j=2}^{i-1}
\sin^2 \theta_j d\,\theta_i^2$ is the element of
solid angle in the $(d-2)-$sphere. For pure Einstein-Hilbert
gravity in vacuum, the solution to the Einstein equations is
\cite{Tangherlini:1963bw}
\be
f(r) = g(r) = 1 - \left(\frac{R_H}{r}\right)^{d-3},
\label{tangher} \ee $R_H$ being the horizon radius. This is the
$d-$dimensional extension of Schwarzschild's solution.

We are interested in extending this solution in the presence of a
dilaton, but considering string-theoretical $\a$ corrections. The effective
action we are thus considering, in the Einstein frame, is given by

\be \label{eef}
\frac{1}{16 \pi G} \int \sqrt{-g} \left( \R - \frac{4}{d-2} \left( \d^\mu \phi \right) \d_\mu \phi + \lambda'\ \mbox{e}^{\frac{4}{d-2} \left( 1 + w \right) \phi} Y(\R) + \mathcal{L}_{\mathrm{matter}}\right) \mbox{d}^dx .
\ee
\noindent
Here, $Y(\R)$ is a scalar polynomial in the Riemann tensor representing the leading higher derivative string corrections to the metric tensor field, and $w$ is its conformal weight, with the convention that $w \left( g_{\mu\nu} \right) = +1$ and $w \left( g^{\mu\nu} \right) = -1$. $\phi$ is the dilaton field, and $\lambda'$ is, up to a numerical factor, the suitable power of the inverse string tension $\a$ for $Y(\R)$. $\mathcal{L}_{\mathrm{matter}}$ contains terms, up to the same order in $\a,$ including other fields than the metric and the dilaton, depending on the type of superstring theory we are considering. In this article we are only considering gravitational $\a$ corrections: therefore we can consistently settle $\mathcal{L}_{\mathrm{matter}}$ to zero.

Defining $T_{\mu\nu}= \frac{\delta Y(\R)}{\delta g^{\mu\nu}}$, the equations of motion for the dilaton and metric obtained from (\ref{eef}) are

\bea
\nabla^2 \phi - \frac{\lambda'}{2}\ Y(\R) = 0, \label{bdfe} \\
\R_{\mu\nu}-\frac{1}{2} g_{\mu\nu} \R = \lambda' T_{\mu\nu} + \frac{1}{2} \lambda' Y(\R) g_{\mu\nu}. \label{bgfe}
\eea

In both equations above we have eliminated certain terms involving powers of $\phi$ (namely, we have omitted the factor $\mbox{e}^{\frac{4}{d-2} \left( 1 + w \right) \phi}$, taking it equal to 1) , since those terms would only contribute at higher orders in our perturbative parameter $\lambda'$.

In this article we are focusing in particular in curvature corrections of order $\a^3$, which are present in general in string theories, and in particular
are the leading corrections in type II superstring effective actions \cite{Gross:1986iv}. In this case, $\lambda'=\frac{\zeta(3)}{16}\ \a^3$, $\zeta(s)$ being the Riemann zeta function, and
\be
Y(\R) = 2 \R_{\mu\nu\rho\sigma} \R_{\lambda\ \ \tau}^{\ \nu\rho} \R^{\mu \eta \theta \lambda} \R_{\ \eta \theta}^{\tau\ \ \sigma} + \R_{\mu\nu\rho\sigma} \R_{\lambda\tau}^{\ \ \rho\sigma} \R^{\mu \eta \theta \lambda} \R_{\ \eta \theta}^{\tau\ \ \nu}. \label{y1}
\ee

We are interested in
computing the first $\lambda'$ corrections to $\phi$ and $g_{\mu\nu},$
using (\ref{bdfe}) and (\ref{bgfe}), taking (\ref{metric1}) with $f(r)=g(r)$ as the $\lambda'=0$ metric and working
perturbatively in $\lambda'.$ In particular we take the $\lambda'=0$ metric in order to
compute $Y(\R)$, since this term is already multiplied by
$\lambda',$ obtaining
\bea
Y(\R) &=& 6(d-4)(d-3)(d-2) \frac{\left(1-f\right)^4}{r^8} + 4 (d-3)(d-2) \frac{\left(1-f\right)^2 f'^2}{r^6} \nonumber \\
&-&2 (d-3)(d-2) \frac{\left(1-f\right) f'^3}{r^5} + \frac{(d-2)^2}{4} \frac{f'^4}{r^4} + (d-2) \frac{f'^3 f''}{r^3} + (d-2) \frac{f'^2 f''^2}{r^2}. \label{y2}
\eea

One also has $\nabla^\mu \nabla_\mu \phi(r) = \left(f
\phi'\right)'+\frac{d-2}{r} f \phi',$ from which we get $r^{d-2}
\nabla^\mu \nabla_\mu \phi(r) = \left( r^{d-2} f \phi' \right)'.$
Putting everything together, replacing $f(r)$ by (\ref{tangher}) and defining
\be
B_d \equiv 2(d-3) (d-2) \left(4d^4 -51 d^3+242 d^2 -489 d +330 \right),
\ee
we write (\ref{bdfe}) as
\be
\left( \left(r^{d-2} -R_H^{d-3} r\right) \phi'\right)'= \frac{\zeta(3)}{16}\ \a^3
\frac{(d-1) B_d}{8} \frac{R_H^{4d-12}}{r^{3d-2}}.
\ee
We simply integrate this equation, obtaining
\be
\left(r^{d-2} -R_H^{d-3} r\right) \phi' = - \frac{\zeta(3)}{16}\ \a^3 \frac{B_d}{24} \frac{R_H^{4d-12}}{r^{3d-3}} - (d-3) \Sigma.
\label{dilinha} \ee The integration constant $\Sigma,$ as will
become clear below, is the dilatonic charge. Integrating again,
defining the incomplete Euler beta function as
$\mathbf{B}(x;\,a,b)=\int_0^x t^{a-1}\,(1-t)^{b-1}\,dt,$ and further defining from now on $z=\frac{\zeta(3)}{16}\ \frac{\a^3}{R_H^6},$ we find \footnote{This solution may also be expressed in terms of the hypergeometric function $_2 F_1$, since the relation $\frac{6}{d-3} \mathbf{B}
\left(\left(\frac{R_H}{r}\right)^{d-3};\, \frac{6}{d-3}, 0
\right) = \left(\frac{R_H}{r}\right)^{\frac{6}{d-3}}\, _2 F_1\left(\frac{6}{d-3},\, 1;\, 1+\frac{6}{d-3}; \, \left(\frac{R_H}{r}\right)^{d-3}\right)$ is valid.}
\bea
\phi(r) &=& -\frac{\Sigma}{R_H^{d-3}} \ln\left( 1 -
\left(\frac{R_H}{r}\right)^{d-3} \right)
-z \frac{B_d}{24}
\left[\frac{1}{6}\left(\frac{R_H}{r}\right)^6 + \frac{1}{d+3}
\left(\frac{R_H}{r}\right)^{d+3} \right. \label{dil} \\
&+& \left. \frac{1}{2d} \left(\frac{R_H}{r}\right)^{2d} + \frac{1}{3d-3} \left(\frac{R_H}{r}\right)^{3d-3}- \frac{1}{d-3} \mathbf{B}
\left(\left(\frac{R_H}{r}\right)^{d-3};\, \frac{6}{d-3}, 0
\right)\right]. \nonumber \eea
At the horizon one only has a coordinate (but not curvature) singularity. From (\ref{bdfe}), this means that also $\phi(r)$ and
$\phi'(r)$ must be nonsingular at $R_H.$ From (\ref{dilinha}) we
see that, in order to avoid $\phi'$ becoming infinite at $r=R_H,$
$\Sigma$ must take a precise value, given by
\be
\Sigma=-\frac{B_d}{24 (d-3)} z R_H^{d-3}. \label{sigma}
\ee
Equation
(\ref{dil}) with $\Sigma$ given by (\ref{sigma}) is the solution
for the dilaton in the background of a spherically symmetric black
hole with $\a^3$
corrections in $d$ dimensions. This dilaton solution acts as
secondary hair, since it does not introduce any new physical
parameter besides the ones of the black hole.

While integrating (\ref{dilinha}), we chose the integration constant so that, at asymptotic infinity, the dilaton vanishes. For large $r$, $\phi$ is approximately given by
\be
\phi(r) \approx \frac{\Sigma}{R_H^{d-3}} \sum_{n=1}^{+\infty} \left(\frac{R_H}{r}\right)^{(d-3)n} +z \frac{B_d}{48} \left[ \frac{1}{2d-3} \left(\frac{R_H}{r}\right)^{2d-3} + \frac{2}{5d-9} \left(\frac{R_H}{r}\right)^{5d-9}+ \ldots\right].
\label{dildeser} \ee
Taking the first term in the series from the logarithm in (\ref{dildeser}), $\phi(r) \approx \frac{\Sigma}{r^{d-3}},$ one can verify that $\Sigma$ has the meaning of a dilatonic charge.

At the horizon, $\phi$ is indeed regular and given by
\footnote{The digamma function is given by
$\psi(z)=\Gamma'(z)/\Gamma(z),$ $\Gamma(z)$ being the usual
$\Gamma$ function. For positive $n,$ one defines
$\psi^{(n)}(z)=d^n\,\psi(z)/d\,z^n.$ This definition can be
extended for other values of $n$ by fractional calculus analytic
continuation. These are meromorphic functions of $z$ with no
branch cut discontinuities.

$\gamma$ is Euler's constant, defined by $\gamma=\lim_{n \to
\infty} \left(\sum_{k=1}^n \frac{1}{k} - \ln n \right),$ with
numerical value $\gamma \approx 0.577216.$}

\be
\phi\left(R_H\right)=-\frac{B_d}{24(d-3)} z \left( \left(\psi^{(0)}\left(\frac{6}{d-3}\right) + \gamma \right) + \frac{(d-3)(d+1) \left(d^2+12d-9 \right)}{6d (d+3)(d-1)}\right). \label{firh}
\ee

From (\ref{dilinha}) and (\ref{sigma}), the derivative of the
dilaton field is given by
$$\phi'\left(r\right)= \frac{B_d}{24} z \frac{R_H^{d-3}}{r^{d-2}}
\frac{1-\left(\frac{R_H}{r}\right)^{3d-3}}{1-\left(\frac{R_H}{r}\right)^{d-3}}.$$
Since $B_d >0$ for $d\geq4,$ we see by inspection that $\phi'\left(r\right)$ is a strictly positive function for $r > R_H;$ we conclude that,
outside the horizon, $\phi$ grows between $\phi\left(R_H\right)$
given by (\ref{firh}) and 0, its value at infinity.

Comparing to the result for the dilaton in the same background but with $\a$ corrections obtained in \cite{Moura:2009it}, one can find the same leading term $-\frac{\Sigma}{R_H^{d-3}} \ln\left( 1 - \left(\frac{R_H}{r}\right)^{d-3} \right)$, obtained in the same way (an integration constant after the first integration of (\ref{bdfe})), but with a different value of the dilaton charge $\Sigma$. It was also found a dependence on the incomplete beta function, but with a different argument: $\textbf{B} \left(\left(\frac{R_H}{r}\right)^{d-3};\, \frac{2}{d-3}, 0 \right)$. The $\a$ corrections considered in \cite{Moura:2009it} were of the same form as (\ref{eef}), but with $Y(\R)$ quadratic in the Riemann tensor, verifying $Y(\R) \propto \left(\frac{R_H}{r}\right)^{2d-2}$. In our case with $\a^3$ corrections we got $Y(\R) \propto \left(\frac{R_H}{r}\right)^{4d-4}$ and a dependence on $\textbf{B} \left(\left(\frac{R_H}{r}\right)^{d-3};\, \frac{6}{d-3}, 0 \right)$. It is logical to conjecture that, with $\left(\a\right)^n$ corrections and in the same background, one should get $Y(\R) \propto \left(\frac{R_H}{r}\right)^{(n-1)(d-1)}$ and a dependence on $\textbf{B} \left(\left(\frac{R_H}{r}\right)^{d-3};\, \frac{2n}{d-3}, 0 \right)$ by the dilaton.
The fraction $\frac{2n}{d-3}$ is an integer for $d=4, 5$ and a half-integer for $d=7$; for these values of $d$ the function $\textbf{B} \left(\left(\frac{R_H}{r}\right)^{d-3};\, \frac{2n}{d-3}, 0 \right)$ can always be written in terms of elementary functions of calculus. For other values of $d$ that may be possible, depending on $n$. In the case $n=3$ we are considering, that is possible for $d= 4, 5, 6, 7, 9$ and 15. For the case $d=15$, the system cannot result from a compactification of a superstring theory, formulated originally in 10 spacetime dimensions. It may result from a compactification of bosonic string theory, although in this case the leading corrections are not of order $\a^3$, but of order $\a$, like those considered in \cite{Moura:2009it}.

\section{The $\a^3-$corrected spherically symmetric black hole}
\label{myersbh}
\noindent

After having obtained the $\a^3-$corrected dilaton, it would be interesting to obtain the $\a^3$ corrections to the black hole solution to which it couples; those would be the leading $\a$ corrections in type II superstring theory for arbitrary $d$.

As we mentioned, $\a$ corrections have first been obtained to first order for generic $d$ in \cite{cmp89}, but these corrections happen to vanish (in the Einstein frame) precisely for $d=4$. This vanishing can be understood from the fact that such $\a$ corrections are given by the Gauss-Bonnet combination $\R^2_{GB}:=\R^{\mu\nu\rho\sigma} \R_{\mu\nu\rho\sigma} - 4 \R^{\mu\nu} \R_{\mu\nu} + \R^2$. In $d=4$ (and in the Einstein frame) this term is topological and therefore it does not contribute to the metric field equations. Therefore, for $d=4$ the $\a^3$ corrections we are considering are the leading corrections to a spherically symmetric black hole even in bosonic and heterotic string theories.

As we saw, a metric including such corrections would be of the form (\ref{metric1}), with

\bea
f(r) &=& \left(1 - \left(\frac{R_H}{r}\right)^{d-3}\right) \left(1+ 2 z\, \mu(r) \right), \nonumber \\
g(r) &=& \left(1 - \left(\frac{R_H}{r}\right)^{d-3}\right) \left(1 -2 z\, \varepsilon(r) \right).
\label{fcgc}
\eea
The factor $1 - \left(\frac{R_H}{r}\right)^{d-3}$ represents the $\a=0$ Tangherlini solution (\ref{tangher}); the functions $\mu(r), \varepsilon(r)$, to be determined, encode the $z$ corrections.

That solution for the lagrangian we are considering was obtained in \cite{Myers:1987qx}, in a
system of coordinates such that the horizon radius $R_H$ is fixed
and has no $\a$ corrections. The result is of the form (\ref{fcgc}), with
\bea
\varepsilon(r)&=& D_d \left(\frac{R_H}{r}\right)^{3d-3}+ E_d \left(\frac{R_H}{r}\right)^{d-3} \frac{1-\left(\frac{R_H}{r}\right)^{2d}}{ 1-\left(\frac{R_H}{r}\right)^{d-3}}, \label{epsilon} \\
\mu(r)&=& -\varepsilon(r) -C_d \left(\frac{R_H}{r}\right)^{3d-3}, \label{mumyers}
\eea
and having defined
\bea
C_d&=& \frac{32}{3}(d-3)(d-1) \left(2 d^3 - 10 d^2 +6d +15\right), \nonumber \\
D_d &=& -\frac{2}{3} (d-3) \left( 52 d^4 -375 d^3 + 758 d^2 -117 d -570\right), \nonumber \\
E_d &=& \frac{2}{3} (d-3) \left( 72 d^5 - 652 d^4 +2079 d^3 - 2654 d^2 + 837 d +570\right).
\label{myers}
\eea
In the same article, the $\a^3-$corrected black hole mass was obtained as
\be
M=\left[1-2 \, z \, E_d \right] \frac{\left(d-2\right) \Omega_{d-2}}{16 \pi G} R_H^{d-3}. \label{massmyers}
\ee

The temperature of a black hole of the form (\ref{fcgc}) is obtained, to first order in $z$, from
\be
T=\lim_{r\rightarrow R_H}\frac{\sqrt{g}}{2\pi} \frac{d\,\sqrt{f}}{d\,r}= \frac{d-3}{4 \pi R_H}\left[1+ z \left(\mu(R_H) - \varepsilon(R_H) \right)\right]. \label{tfg}
\ee
In our case, the result is given by
\be
T = \frac{d-3}{4 \pi R_H}\left[1- z F_d \right], \, F_d=C_d + 2 D_d + \frac{4d}{d-3} E_d. \label{temp}
\ee

The black hole entropy for this solution wasn't studied in \cite{Myers:1987qx}, but it can be obtained using Wald's
formula \cite{w93}
\be
S=-2 \pi \int_H \frac{\partial
\mathcal{L}}{\partial \R^{\mu \nu \rho \sigma}}
\varepsilon^{\mu\nu} \varepsilon^{\rho\sigma} \, \sqrt{h} \,
d\,\Omega_{d-2}, \label{wald}
\ee
where $H$ is the black hole horizon, with area $A_H=R_H^{d-2} \Omega_{d-2}$ and metric
$h_{ij}$ induced by the spacetime metric $g_{\mu\nu}.$ For the
metric (\ref{metric1}), the nonzero components of the binormal
$\varepsilon^{\mu\nu}$ to $H$ are
$\varepsilon^{tr}=-\varepsilon^{rt}=-\sqrt{\frac{g}{f}}.$ From the $\lambda'-$corrected effective action (\ref{eef}) one also needs
$$8 \pi G\frac{\partial \mathcal{L}}{\partial \R^{\mu \nu \rho \sigma}}=\frac{1}{4}\left(g_{\mu\rho}g_{\sigma\nu}-g_{\mu\sigma}g_{\rho\nu}\right)+ \mbox{e}^{\frac{4}{d-2} \left( 1 + w \right) \phi} \lambda' \frac{\partial Y(\R)}{\partial \R^{\mu \nu \rho \sigma}}.$$

Taking only nonvanishing components from the equation above and considering now the $\a^3$ correction (\ref{y1}) with conformal weight $w=-4$, one gets $$8 \pi G \frac{\partial \mathcal{L}}{\partial
\R^{\mu \nu \rho \sigma}} \varepsilon^{\mu\nu}
\varepsilon^{\rho\sigma} = 4 \times 8 \pi G \frac{\partial
\mathcal{L}}{\partial \R^{trtr}} \varepsilon^{tr} \varepsilon^{tr}
=\left(-\frac{f}{g}+\mbox{e}^{-\frac{12}{d-2} \phi} \lambda'
\frac{2(d-2)(f')^2 \left(f'+ 2r f''\right) }{r^3}\right)\frac{g}{f}.$$ The $\lambda'-$corrected term must be evaluated at order $\lambda'=0$. At this order $\phi=0,$ $f=g$ (given by (\ref{tangher})) and $2(d-2)(f')^2 \left(f'+ 2r f''\right)=-\frac{2(d-2)(d-3)^2 (2d-5)}{R_H^3} \left(\frac{R_H}{r}\right)^{3d-6}$. Taking this term evaluated at the horizon and replacing $\lambda'$ by $z$, one therefore gets for the $\a^3-$corrected entropy
\be
S=\frac{1}{4 G} \int_H \left( 1 + 2(d-2)(d-3)^2 (2d-5)z\right) \, \sqrt{h} \, d\,\Omega_{d-2} =\frac{A_H}{4 G}
\left(1+ 2(d-2)(d-3)^2 (2d-5) z\right). \label{pmes}
\ee

One can obtain the black hole free energy through the relation $F=M-TS:$
\bea
F&=&\left[1+ z \, G_d \right] \frac{\Omega_{d-2}}{16 \pi G} R_H^{d-3}, \label{free} \\
G_d&=&\frac{2}{3}(d-3)\left(144 d^6- 1232 d^5 + 3622 d^4 - 4021 d^3 + 278 d^2 + 3003 d-1830\right).
\nonumber
\eea

The black hole specific heat is given by $C= T
\frac{\partial\,S}{\partial\,T}.$ Since for the expressions we obtained only the horizon radius is a variable, we may fully express $T$ and $S$ as functions of $R_H$ and vice-versa (including the contribution from $z=\frac{\lambda'}{R_H^6}$). We then have
\be
C=T \frac{\frac{d\,S}{d\,R_H}}{\frac{d\,T}{d\,R_H}}. \label{cc}
\ee
In our case, we get
\be
C= -(d-2)\frac{A_H}{4G} \left[1+z \, H_d \right], \, H_d= 2(2d-5)(d-3)^2 (d-8) + 6 F_d. \label{ccmyers}
\ee

It is useful to express the physical quantities we have been computing in a covariant way, in terms of global charges of the black hole, and not in terms of quantities that may depend on the system of coordinates. For the solution we are considering, the most obvious choice is to express those quantities in terms of the black hole mass $M$. In order to do this, we invert (\ref{massmyers}), obtaining to first order in the perturbative parameter $z$ (which we also express in terms of $M$)
\bea
R_H&=&\left(\frac{16\, \pi\, G\, M}{\left(d-2\right) \Omega_{d-2}}\right)^{\frac{1}{d-3}} \left[1+\frac{2}{d-3} \, E_d z \right], \label{rhm} \\
z&=&\frac{\zeta(3)}{16}\ \left(\frac{\left(d-2\right) \Omega_{d-2}}{16\, \pi\, G\, M}\right)^{\frac{6}{d-3}} \a^3. \label{zm}
\eea
Expression (\ref{rhm}) must be interpreted with care. While solving the $\a^3-$corrected field equations, we took a system of coordinates such that the horizon radius $R_H$ had no $\a^3$ corrections. This way, we obtained $\a^3$ corrections for the mass, given (in such system of coordinates) by (\ref{massmyers}). Now, we are choosing the black hole mass $M$ as the parameter having no $\a^3$ corrections. This is equivalent to $R_H$ having such corrections, given in leading order by (\ref{rhm}). Those corrections are given as a perturbative series in the parameter $z$, expressed now in terms of $M$ by (\ref{zm}). To obtain the $\a^3$ corrections in terms of the mass to the thermodynamical quantities we previously considered, we simply replace (\ref{rhm}) in the expressions (\ref{temp}), (\ref{pmes}), (\ref{free}) and (\ref{ccmyers}), respectively, and expand each expression to order $z$, obtaining
\bea
T &=& \frac{d-3}{4 \pi} \left(\frac{\left(d-2\right) \Omega_{d-2}}{16\, \pi\, G\, M}\right)^{\frac{1}{d-3}} \left[1- \left(F_d +\frac{2}{d-3} \, E_d\right) z \right], \label{tempm} \\
S&=&\frac{1}{4 G \Omega_{d-2}^{\frac{1}{d-3}}} \left(\frac{16\, \pi\, G\, M}{d-2}\right)^{\frac{d-2}{d-3}}
\left[1+ \left(2(d-2)(d-3)^2 (2d-5) +2 \frac{d-2}{d-3} \, E_d \right) z \right], \label{pmesm} \\
F&=& \frac{M}{d-2} \left[1+ (G_d + 2 E_d) z \right], \label{freem} \\
C&=& -\frac{1}{4 G} \frac{\left(16\, \pi\, G\, M\right)^{\frac{d-2}{d-3}}}{\left((d-2) \Omega_{d-2}\right)^{\frac{1}{d-3}}} \left[1+\left(H_d +2 \frac{d-2}{d-3} \, E_d\right) z \right]. \label{ccmyersm}
\eea
These $\a^3$ corrections are the leading $\a$ corrections to these quantities in type II superstring theory (and, for $d=4$, also in heterotic and bosonic string theory), expressed covariantly in terms of the black hole mass.

We have checked that, for every relevant value of $d$, $E_d>0$, $F_d>0$ and $G_d>0$. Therefore, we conclude that the $\a^3$ corrections to the black hole entropy and free energy are positive. We have also checked that, for every relevant value of $d$, $H_d +2 \frac{d-2}{d-3} \, E_d >0$, from which we conclude that in the presence of $\a^3$ corrections this black hole keeps being thermodynamically unstable, with a negative specific heat.

We can also conclude from this analysis that the $\a^3$ corrections to the black hole temperature are negative. One expects the $\a$ corrections (specially of order $\a^3$, as we are considering) to be smaller in magnitude than the $\a=0$ terms, as it is usual in perturbation theory. Nonetheless, one should check if, just considering the leading $\a^3$ correction, one does not get a negative temperature. From (\ref{tempm}), this means having $1- \left(F_d +\frac{2}{d-3} \, E_d\right) z>0$, or
\be
\frac{\a}{\left(G\, M\right)^{\frac{2}{d-3}}}<\sqrt[3]{\frac{16}{\zeta(3) \left(F_d +\frac{2}{d-3} \, E_d\right)}} \left(\frac{16\, \pi}{\left(d-2\right)\Omega_{d-2}}\right)^{\frac{2}{d-3}}. \label{tm0}
\ee
The power of the black hole mass $M$ in (\ref{tm0}) decreases with the spacetime dimension $d$ but, because of the expected large values of this mass, we do not expect this condition to be very restrictive, even for the largest values of $d$: for $d=10$, condition (\ref{tm0}) implies $\a < 0.00377426 (G M)^{2/7}$. But (\ref{tempm}) is only a first-order perturbative approximation; a complete analysis would require a full knowledge of $T$ to all orders in $\a$. Nonetheless, the leading string correction being negative suggests that there may exist a value of $M$ for which the temperature reaches a maximum. For each particular given value of $d$, the $\a^3$ corrected temperature (\ref{tempm}) has a maximum for $M=\frac{\left(d-2\right) \Omega_{d-2}}{16\, \pi\, G} \left(7\,\frac{\zeta(3)}{16} \, \left(F_d +\frac{2}{d-3} \, E_d\right)\,\a^3\right)^{\frac{d-3}{6}}$, given by
\be
T_{max}=\frac{3}{14 \pi} \frac{d-3}{\left(7\,\frac{\zeta(3)}{16} \, \left(F_d +\frac{2}{d-3} \, E_d\right)\,\a^3\right)^{\frac{1}{6}}}. \label{tmx}
\ee
The numerical values one gets vary between $T_{max}\approx\frac{0.0155}{\sqrt{\a}}$ (for $d=4$) and $T_{max}\approx\frac{0.0265}{\sqrt{\a}}$ (for $d=10$). These values are smaller than the critical Hagedorn temperature obtained from the free spectrum of the superstring, given by $T_{crit}=\frac{1}{\pi \sqrt{8 \a}}\approx\frac{0.11}{\sqrt{\a}}$ \cite{Myers:1987qx,Sundborg:1984uk,Bowick:1985az}.

\section{The $\a^3-$corrected black hole with a toroidal compactification}
\label{compbh}
\noindent

Article \cite{Myers:1987qx} considers black holes on arbitrary $d$ spacetime
dimensions in the presence of a dilaton and string $\a^3$ corrections. Since string theories are formulated in $D$ spacetime
dimensions ($D=26$ for bosonic strings and $D=10$ for superstrings), one should consider the effects of compactification from 10
or 26 to $d$ dimensions.

When one talks about a black hole in string theory in $d$ dimensions, the original
$D$--dimensional spacetime must have been compactified on some
$(D-d)$--dimensional manifold, with internal coordinates $y^m$
and internal metric $g_{mn}(y).$ When passing from the string to
the Einstein frame, one needs a transformation under which
\be
g_{\mu\nu} \rightarrow \exp \left( \frac{4}{d-2} \Phi \right)
g_{\mu\nu}, \,\, {\R_{\mu\nu}}^{\rho\sigma} \rightarrow
{\widetilde{\R}_{\mu\nu}}^{\ \ \ \rho\sigma} =
{\R_{\mu\nu}}^{\rho\sigma} -
{\delta_{\left[\mu\right.}}^{\left[\rho\right.} \nabla_{\left.\nu
\right]} \nabla^{\left.\sigma \right]} \Phi. \label{sf} \ee If one
takes this as a conformal transformation of the entire
$D-$dimensional metric (rather than just on the $d-$dimensional
black hole part, as it was done in \cite{Myers:1987qx}), it involves the total dilaton field $\Phi,$
including the Kaluza-Klein part depending on the internal
coordinates $y^m$ (rather than just the $d-$dimensional part
$\phi$ as we have been considering). This way the size of the
compact space becomes spatially varying, being governed by a
function $h.$ Since, for the cases we have been considering, the dilaton field depends only on the radial coordinate $r$, the same is to be expected for the function $h$. The complete line element is then the sum of the $d-$dimensional black hole (\ref{metric1}) and the compact space:
\be
d\,s^2=-f(r)\,d\,t^2 + g^{-1}(r)\,d\,r^2+r^2\,d\,\Omega^2_{d-2} +
h \, g_{mn}(y)\, d\,y^m\,d\,y^n. \label{metric2} \ee
This metric is a solution of the metric field equation (\ref{bgfe}) for the whole spacetime, in $D$ dimensions, which we write as
\be
\R_{\mu\nu} + \lambda'\ \left( \frac{1}{D-2} Y(\R) g_{\mu\nu} + \frac{1}{D-2} g_{\mu\nu} T^\rho_{\,\,\rho} - T_{\mu\nu} \right) = 0. \label{4gfe}
\ee
From (\ref{4gfe}), the compact space and the black hole cannot be decoupled in general: the respective curvatures appear combined on the terms depending on $Y(\R)$ and $T_{\mu\nu}$. In order to avoid this problem, we take the internal
space to be a flat torus, with vanishing internal curvature. Also since, for the cases we have been considering, the dilaton field depends only on the radial coordinate $r$, the same is to be expected for the function $h$. At order $\lambda'=0$, $\phi=0$ and the conformal transformation (\ref{sf}) is just the identity. This means one should then have
\be
g_{mn}(y)=\delta_{mn},\,\,h(r)=1+2 \lambda' \rho(r). \label{hd}
\ee
One must now determine the function $\rho(r)$. By contracting (\ref{4gfe}) with the $D-$dimensional metric, one finds the $D-$dimensional Ricci scalar $\R^D=\sum_{\mu, \nu =1}^D \R_{\mu\nu} g^{\mu\nu}$:
\be
\R^D + \frac{\lambda'}{D-2} \left(D\ Y(\R)  + 2\ T^\rho_{\,\,\rho}\right) =0. \label{RD}
\ee
But if one rather contracts (\ref{4gfe}) with just the $d-$dimensional part of the metric (\ref{metric2}), i.e. the black hole metric (\ref{metric1}), one obtains $\R^d=\sum_{\mu, \nu =1}^d \R_{\mu\nu} g^{\mu\nu}$:
\be
\R^d + \frac{\lambda'}{D-2} \left(d\ Y(\R) -(D- 2-d)\ T^\rho_{\,\,\rho}\right) =0. \label{rd}
\ee
Equations (\ref{RD}) and (\ref{rd}) were obtained from the field equation (\ref{4gfe}). But one can take directly the $D-$dimensional metric (\ref{metric2}), with the specifications (\ref{hd}), and compute its corresponding Ricci tensor $\R_{\mu\nu}$. One can then contract it with the whole metric, obtaining the Ricci scalar $\R^D$, or just with the $d-$dimensional black hole part, obtaining $\R^d$. Proceeding this way, one verifies that
\be
\R^D-\R^d=-(D-d) \lambda' \nabla^2 \rho. \label{Rd}
\ee
Combining (\ref{RD}), (\ref{rd}) and (\ref{Rd}), we conclude that $\rho$ must satisfy the equation
\be
\lambda' \nabla^2 \rho= \frac{\lambda'}{D-2} \left(Y(\R) +\ T^\rho_{\,\,\rho}\right). \label{rho}
\ee
$Y(\R)$ and $T^\rho_{\,\,\rho}$ should be evaluated with the $\lambda'=0$ metric (\ref{metric1}), with $f=g$ given by (\ref{tangher}).

For the case of $\a^3$ corrections we are dealing with, $Y(\R)$ is given by (\ref{y1}) or (\ref{y2}), before or after replacing the metric. The explicit expression for $T_{\mu\nu}$ can be found in article \cite{Grisaru:1986px}; we choose not to repeat it here, since it is very long and all that we need is the result for $T^\rho_{\,\,\rho}.$

After integrating (\ref{rho}) and requiring $\rho$ to be finite at the horizon (a similar procedure to the one taken to obtain (\ref{dilinha})), we are left with
\bea
&&\left(r^{d-2} -R_H^{d-3} r\right) \rho' = -\frac{2\  R_H^{d-9}}{D-2}\ \left( \left(A_d-B_d\right)\left(\frac{R_H}{r}\right)^{2d} -A_d \left(\frac{R_H}{r}\right)^{3d-3} +B_d \right), \label{rholinha} \\
&& A_d=2 (d - 3) (d - 2)^2 (8 d^4 - 80 d^3 + 233 d^2 - 152 d - 45).
\eea
Equation (\ref{rholinha}) can be integrated and conveniently multiplied by $z$, in order to finally obtain
\be
z \rho(r) =\frac{2}{D-2} \left[ z\frac{A_d-B_d}{3(d-1)} \left(\frac{R_H}{r}\right)^{3d-3}- 24 \phi(r)\right], \label{rhof}
\ee
with $\phi(r)$ given by (\ref{dil}).

We now proceed to determine the influence of the internal compact space (the torus) on the $d-$dimensional black hole geometry. There are two nontrivial components of the field equation (\ref{4gfe}), corresponding to $\R_{tt}$ and $\R_{rr}$ \cite{cmp89,Myers:1987qx}. We use these two equations in order to obtain the two unknown functions $\mu(r), \, \varepsilon(r)$ in (\ref{fcgc}). For $\varepsilon(r)$ we obtain the same result as (\ref{epsilon}), while $\mu(r)$ is now given by
\bea
\mu(r)= -\varepsilon(r) -C_d \left(\frac{R_H}{r}\right)^{3d-3}-\Delta(d,r) \label{muc}, \\
\Delta(d,r)=\frac{D-d}{D-2} \left[\rho(r)-r \rho'(r)\right]. \label{del}
\eea
Metric (\ref{metric1}), with $f(r), g(r)$ of the form (\ref{fcgc}), $\varepsilon(r)$ given by (\ref{epsilon}) and $\mu(r)$ given by (\ref{muc}), corresponds to the $d-$dimensional black hole solution in the presence of a $(D-d)-$dimensional compact torus we have been looking for.

Although in most cases we will have $D=10$ as the dimension of the original spacetime (the critical dimension of superstring theory), as we discussed in the previous section, for a spherically symmetric metric in $d=4$ and in the Einstein frame $\a^3$ corrections are in general the leading corrections, like we are considering. This is true also in bosonic string theory, whose critical dimension is $D=26$. Because of this possibility we chose to leave the value of $D$ unspecified in our solution.

\section{Thermodynamical properties of the compactified black hole}
\noindent

In this section, we compute several thermodynamical quantities for
the black hole solution we have just found. In each case we
compare the result to the corresponding one of the noncompactified solution from \cite{Myers:1987qx}
obtained in section \ref{myersbh}, since the parameters are the
same. This way we can evaluate the effects on the physical quantities introduced by
the toroidal compactification. These effects, as we will see, are all expressed in terms of the function $\Delta(d,r)$ given by (\ref{del}). This function vanishes equally for $d=D$, when no compactification is present. Every result that is obtained in this section matches the corresponding one obtained in section \ref{myersbh} if one sets $\Delta(d,r) \equiv 0$.

The entropy of this black hole solution can be obtained by Wald's
formula (\ref{wald}). It is clear from this formula that the $\lambda'$--correction to the entropy depends only
on the $\lambda'=0$ part of the metric. Since this part of the metric is the same for the cases we considered, the
result for the entropy does not change: it is given by (\ref{pmes}).

The free energy of a black hole solution is
obtained from the euclideanized action (\ref{eef}),
to which one adds a surface term consisting of an integral (on the
boundary) of the trace of the second fundamental form, subtracted
by the same trace for the boundary embedded on flat space, to
render the total surface contribution finite \cite{Gibbons:1976ue}. This surface term
also includes contributions for the higher-derivative terms, but
these contributions do not affect this calculation \cite{cmp89}.
Because we chose a system of coordinates such that $R_H$ does not get $\a$ corrections, there are no implicit $\a$ corrections: all the $\lambda'$--correction terms in the euclidean action are explicit and should be evaluated using the $\lambda'=0$ part of the metric. This means that, just as it happened for the entropy, in the system of coordinates where $R_H$ does not get $\a$ corrections (and  just in this system of coordinates) the result for the free energy for our solution is the same as that for the noncompactified solution obtained in \cite{Myers:1987qx}, given by (\ref{free}).

The black hole temperature is given by (\ref{tfg}). Using (\ref{epsilon}) and (\ref{muc}), but also (\ref{firh}), (\ref{rholinha}), (\ref{rhof}), we obtain
\be
T = \frac{d-3}{4 \pi R_H}\left[1- z \left(F_d +\Delta(d,R_H) \right)\right]. \label{tempt}
\ee
$\Delta(d,R_H)$ is the function (\ref{del}) evaluated at the horizon, given by
$$
\Delta(d,R_H)= \frac{D-d}{3\left(D-2\right)^2} \left[\left[\frac{6}{d-3} \left(\psi^{(0)}\left(\frac{6}{d-3}\right) + \gamma \right) + \frac{d^2+12d+9 }{d (d+3)} + \frac{12d}{d-3}\right]B_d 
+
\left(6 +\frac{2}{d-1}\right) A_d\right].
$$

The black hole inertial mass is given by $M_I=\frac{\left( d-2 \right) \Omega_{d-2}}{16 \pi G} \lim_{r \to \infty} r^{d-3} \Big(
1 - g \left(r\right) \Big)$, while the gravitational mass is given by $M_G=\frac{\left( d-2
\right) \Omega_{d-2}}{16 \pi G} \lim_{r \to \infty} r^{d-3} \Big(1 - f \left(r\right) \Big)$. Since from (\ref{fcgc}), (\ref{muc}), (\ref{del}) and also (\ref{rhof}) $f(r)-g(r)$ is of order $\frac{1}{r^{d-3}}$, one finds indeed $M_G\neq M_I$. This situation is usual when one is dealing with compactifications and originates from the integration of Kaluza-Klein modes in the full $D-$dimensional action, resulting in a $d-$dimensional action with nondiagonal kinetic terms.

The actual physical mass can be obtained from the relation $M=ST+F.$ From (\ref{pmes}), (\ref{free}), (\ref{tempt}),
\be
M=\left[1-z \left(2 E_d+\frac{d-3}{d-2} \Delta(d,R_H) \right)\right] \frac{\left(d-2\right) \Omega_{d-2}}{16 \pi G} R_H^{d-3}. \label{pmem}
\ee

The black hole specific heat is given by (\ref{cc}). In our case, we get
\be
C= -(d-2)\frac{A_H}{4G} \left[1+z \left(H_d + 6 \Delta(d,R_H)\right) \right].
\ee

By inverting (\ref{pmem}), like we did in section \ref{myersbh} with (\ref{massmyers}), we can now express these thermodynamical quantities for this solution in terms of the black hole mass $M$:
\bea
T &=& \frac{d-3}{4 \pi} \left(\frac{\left(d-2\right) \Omega_{d-2}}{16\, \pi\, G\, M}\right)^{\frac{1}{d-3}} \left[1- \left(F_d +\frac{2}{d-3} \, E_d +\frac{d-1}{d-2} \Delta(d,R_H)\right) z \right], \label{tempt2} \\
S&=&\frac{1}{4 G \Omega_{d-2}^{\frac{1}{d-3}}} \left(\frac{16\, \pi\, G\, M}{d-2}\right)^{\frac{d-2}{d-3}}
\left[1+ \left(2(d-2)(d-3)^2 (2d-5) +2 \frac{d-2}{d-3} \, E_d + \Delta(d,R_H))\right) z \right], \label{pmest} \\
F&=& \frac{M}{d-2} \left[1+ \left(G_d + 2 E_d +\frac{d-3}{d-2} \Delta(d,R_H)\right) z \right], \label{freet} \\
C&=& -\frac{1}{4 G} \frac{\left(16\, \pi\, G\, M\right)^{\frac{d-2}{d-3}}}{\left((d-2) \Omega_{d-2}\right)^{\frac{1}{d-3}}} \left[1+\left(H_d +2 \frac{d-2}{d-3} \, E_d + 7\Delta(d,R_H))\right) z \right]. \label{ccmyerst}
\eea
We have checked that, for every relevant value of $d$, $\Delta(d,R_H)>0$. Therefore, we conclude that also for this solution the $\a^3$ corrections to the black hole temperature are negative, while those corrections to the black hole entropy and free energy are positive. Also the $\a^3$ corrections to the specific heat are such that this black hole keeps being thermodinamically unstable, with $C<0$. The effect of the compactification (and of the presence of $\Delta(d,R_H)$) on these quantities is to increase the magnitude of their $\a^3$ corrections as compared to those of the noncompactified solution of \cite{Myers:1987qx}.

An analysis similar to the one made at the end of section \ref{myersbh} tells us that, in order for the temperature to remain positive in the presence of the leading $\a^3$ correction, (\ref{tm0}) must be changed to
\be
\frac{\a}{\left(G\, M\right)^{\frac{2}{d-3}}}<\sqrt[3]{\frac{16}{\zeta(3) \left(F_d +\frac{2}{d-3} \, E_d+\frac{d-1}{d-2} \Delta(d,R_H)\right)}} \left(\frac{16\, \pi}{\left(d-2\right)\Omega_{d-2}}\right)^{\frac{2}{d-3}}. \label{tmy}
\ee
Once again, the leading string correction being negative suggests that there exists a value of $M$ for which the temperature reaches a maximum, given by an analogous change in (\ref{tmx}):
\be
T_{max}=\frac{3}{14 \pi} \frac{d-3}{\left(7\,\frac{\zeta(3)}{16} \, \left(F_d +\frac{2}{d-3} \, E_d + \frac{d-1}{d-2} \Delta(d,R_H) \right)\,\a^3\right)^{\frac{1}{6}}}. \label{tmc}
\ee
We considered the numerical values of these expressions for every relevant value of $d$, like we did at the end of section \ref{myersbh}, taking $D=10$ when evaluating  $\Delta(d,R_H)$ (i.e. considering now specifically type II superstring compactifications, having (\ref{y2}) as leading $\a^3$ curvature corrections). The only difference between these expressions and (\ref{tm0}) and (\ref{tmx}) is the inclusion of $\Delta(d,R_H)$. We verified that the numerical effects of the inclusion of such term on (\ref{tmy}) and (\ref{tmc}) are very small, affecting typically only the fourth nonzero decimal digits when compared to the results using (\ref{tm0}) and (\ref{tmx}) obtained at the end of section \ref{myersbh}. The conclusions obtained there on the bounds for $\frac{\a}{\left(G\, M\right)^{\frac{2}{d-3}}}$ and for $T_{max}$, and its comparison to the superstring Hagedorn temperature, remain valid for the solution we obtained in section \ref{compbh}.

\section{Conclusions and future directions}
\noindent

In this work, we derived the solution to a dilaton in the presence of a spherically symmetric black hole in string theory with $\a^3$ gravitational corrections in $d$ spacetime dimensions. We then obtained a spherically symmetric black hole solution with $\a^3$ corrections from toroidally compactified superstring theory in the same $d$ dimensions. This is a modification to the solution from article \cite{Myers:1987qx}, which has the same type of $\a^3$ corrections but was obtained directly in $d$ dimensions, ignoring the effects of superstring compactification.

For the solution we obtained, we computed its free energy, entropy, temperature, specific heat and mass. We compared the magnitude of the $\a$ corrections to these thermodynamical quantities to the ones corresponding to the noncompactified solution obtained in \cite{Myers:1987qx}, in order to estimate the effects of string compactification. For both solutions, when these quantities are expressed in terms of the black hole mass, the $\a^3$ corrections to the free energy and entropy are positive; the $\a^3$ corrections to the temperature, on the other hand, are negative. For all such quantities, the magnitude of these $\a^3$ corrections is in general larger for the compactified solution we obtained, when compared to those of the noncompactified one. Both solutions are thermodinamically unstable, like the classical Tangherlini black hole. Understanding the black hole temperature as a function of the mass, for both solutions we estimated its maximum and we verified that such maximum value is smaller than the superstring Hagedorn temperature.

Overall we conclude that neither the $\a^3$ corrections nor the effects of toroidal compactification do qualitatively change the thermodynamical properties of these black holes.

In future works we plan to study some other features of these $\a^3$-corrected black holes, like its stability under perturbations of the metric and quasinormal modes. In a previous work we have made such studies for a similar black hole solution with $\a$ corrections, considering tensorial perturbations of the metric \cite{Moura:2012fq}. This work can be extended to these $\a^3$-corrected black hole solutions we considered and obtained in this article. One can also consider for these studies other kinds of metric perturbations (vector and scalar).

The usual Tolman-Oppenheimer-Volkoff (TOV) equations, describing the hydrostatic equilibrium of compact objects, are obtained in Einstein gravity.
Modified TOV equations have been obtained in the literature for theories including higher derivative corrections in the form of powers of the Ricci scalar \cite{Astashenok:2014nua} and of the Gauss-Bonnet combination \cite{Momeni:2014bua}. None of these approaches considers the effect of the dilaton field; furthermore, the higher derivative corrections that have been considered are not in the form suggested by string theories. The string corrections to the TOV equations may therefore be regarded as an open problem.

The $\a^3$-corrected field equations and the solutions that we considered in this article can be applied to obtain the string corrected TOV equations, also working perturbatively in $\a^3$ as we did, since as we saw they represent the leading string gravitational corrections to spherically symmetric solutions in $d=4$ and in the Einstein frame. The knowledge of these corrections might be very relevant in the strong gravity regime.

\paragraph{Acknowledgements}
\noindent
This work has been supported by Funda\c c\~ao para a Ci\^encia e a Tecnologia under contract IT (UID/EEA/50008/2019).

\end{document}